      [1999/12/01 v1.4c Il Nuovo Cimento]
\documentclass{cimento}
\usepackage{graphics}
\usepackage{graphicx}
\usepackage{epsfig}
\usepackage{amsmath}
\usepackage{bm}

\title{Radiative Corrections to the PREX and QWEAK Experiments}
\author{C. J. Horowitz\from{ins:x}}
\instlist{\inst{ins:x} Center for Exploration of Energy and Matter and Physics Department, Indiana University, Bloomington, IN 47405, USA}
\PACSes{\PACSit{21.10.Gv}{Nucleon distributions and halo features}\PACSit{24.80.+y}{Nuclear tests of fundamental interactions and symmetries}}
\begin{document}

\maketitle

\begin{abstract}
The high precision parity violating electron scattering experiments PREX and QWEAK are sensitive to radiative corrections.  In this paper we introduce the PREX experiment that measures the neutron radius of $^{208}$Pb and discuss coulomb distortion corrections.   We then discuss dispersion corrections for QWEAK that aims to measure the weak charge of the proton.   
\end{abstract}
The experiments PREX \cite{prex} and QWEAK \cite{qweak} aim to measure the parity violating asymmetry in electron scattering with high precision.  This precision necessitates a careful consideration of radiative corrections.  In Sec. \ref{Sec1}, we introduce the PREX experiment that measures the neutron radius of $^{208}$Pb.  Next in Sec. \ref{Sec2}, we discuss radiative corrections, from coulomb distortions for PREX, and dispersion corrections for QWEAK.  

\section{The PREX experiment, neutron densities, and neutron radii}  
\label{Sec1}

Nuclear charge densities have been accurately measured with electron scattering and have become our picture of the atomic nucleus, see for example ref. \cite{chargeden}.  These measurements have had an enormous impact.  
In contrast, our knowledge of neutron densities comes primarily from hadron scattering experiments involving for example pions \cite{pions}, protons \cite{protons1,protons2,protons3}, or antiprotons \cite{antiprotons1,antiprotons2}.  However, the interpretation of hadron scattering experiments is model dependent because of uncertainties in the strong interactions.

Parity violating electron scattering provides a model independent probe of neutron densities that is free from most strong interaction uncertainties.  This is because the weak charge of a neutron is much larger than that of a proton \cite{dds}.  Therefore the $Z^0$ boson couples primarily to neutrons.  In Born approximation, the parity violating asymmetry $A_{pv}$, the fractional difference in cross sections for positive and negative helicity electrons, is proportional to the weak form factor.  This is very close to the Fourier transform of the neutron density.  Therefore the neutron density can be extracted from an electro-weak measurement \cite{dds}.  
Many details of a practical parity violating experiment to measure neutron densities have been discussed in a long paper \cite{bigprex}.    

The doubly magic nucleus $^{208}$Pb has 44 more neutrons than protons, and some of these extra neutrons are expected to be found in the surface where they form a neutron rich skin.
The thickness of this skin is sensitive to nuclear dynamics and provides fundamental nuclear structure information.  There may be a useful analogy with cold atoms in laboratory traps were similar ``spin skins'' have been observed for partially polarized systems  \cite{coldatomsMIT, coldatomsRICE}.  Note that there is an attractive interaction between two atoms of unlike spins while for a nucleus, the interaction between two nucleons of unlike isospins is also more attractive than the interaction between two nucleons of like isospins.       


The neutron radius of $^{208}$Pb, $R_n$, has important implications for astrophysics.  There is a strong correlation between $R_n$ and the pressure of neutron matter $P$ at densities near 0.1 fm$^{-3}$ (about 2/3 of nuclear density) \cite{alexbrown}.  A larger $P$ will push neutrons out against surface tension and increase $R_n$.  Therefore measuring $R_n$ constrains the equation of state (EOS) --- pressure as a function of density --- of neutron matter.  

Recently Hebeler et al. \cite{hebeler} used chiral perturbation theory to calculate the EOS of neutron matter including important contributions from very interesting three neutron forces.  
From their EOS, they predict $R_n-R_p= 0.17 \pm 0.03$ fm.  Here $R_p$ is the known proton radius of $^{208}$Pb.   Monte Carlo calculations by Carlson et al. also find sensitivity to three neutron forces \cite{MC3n}.   Therefore, measuring $R_n$ provides an important check of fundamental neutron matter calculations, and constrains three neutron forces.

The correlation between $R_n$ and the radius of a neutron star $r_{NS}$ is also very interesting \cite{rNSvsRn}.  In general, a larger $R_n$ implies a stiffer EOS, with a larger pressure, that will also suggest $r_{NS}$ is larger.  Note that this correlation is between objects that differ in size by 18 orders of magnitude from $R_n\approx 5.5$ fm to $r_{NS}\approx 10$ km.   

The EOS of neutron matter is closely related to the symmetry energy $S$.  
This describes how the energy of nuclear matter rises as one goes away from equal numbers of neutrons and protons.  There is a strong correlation between $R_n$ and the density dependence of the symmetry energy $dS/dn$, with $n$ the baryon density.  The symmetry energy can be probed in heavy ion collisions \cite{isospindif}.  For example, $dS/dn$ has been extracted from isospin diffusion data \cite{isospindif2} using a transport model.

The symmetry energy $S$ helps determine the composition of a neutron star.     A large $S$, at high density, implies a large proton fraction $Y_p$ that will allow the direct URCA process of rapid neutrino cooling.  If $R_n-R_p$ is large, it is likely that massive neutron stars will cool quickly by direct URCA  \cite{URCA}.  In addition, the transition density from solid neutron star crust to the liquid interior is strongly correlated with $R_n-R_p$ \cite{cjhjp_prl}.  

Recently, Reinhard and Nazarewicz claim that $R_n-R_p$ is strongly correlated with the dipole polarizability $\alpha_D$ of $^{208}$Pb \cite{Reinhard} and Tamii et al. have accurately determined $\alpha_D$ from very small angle proton scattering \cite{Tamii}.   However, further work suggests the correlation of $\alpha_D$ with $R_n-R_p$ is model dependent \cite{JP2011}.

Finally, atomic parity violation (APV) is sensitive to $R_n$ \cite{pollockAPV},\cite{brownAPV},\cite{bigprex}.  Parity violation involves the overlap of atomic  electrons with the weak charge of the nucleus, and this is primarily carried by the neutrons.  Furthermore, because of relativistic effects the electronic wave function can vary rapidly over the nucleus.  Therefore, the APV signal depends on where the neutrons are and on $R_n$.   A future low energy test of the standard model may involve the combination of a precise APV experiment along with PV electron scattering to constrain $R_n$.  Alternatively, measuring APV for a range of isotopes can provide information on neutron densities \cite{berkeleyAPV}.

The Lead Radius Experiment (PREX) measures the parity violating asymmetry $A_{pv}$ for 1.05 GeV electrons scattering from $^{208}$Pb at five degrees \cite{prex}.    This measurement  initially ran at Jefferson Laboratory in the spring of 2010.  PREX demonstrates this new technique to measure neutron densities and achieved the desired small systematic errors.  Therefore the measurement can be improved by accumulating more statistics.  The collaboration purposes to design appropriate engineering modifications to the beamline to mitigate radiation problems and has been granted additional beam time to improve the statistics and achieve the original goal of a 1\% ($\pm 0.05$ fm) constraint on the neutron radius of $^{208}$Pb.

In addition to PREX, many other parity violating measurements of neutron densities are possible, see for example \cite{PREXII}.  Neutron radii can be measured in many stable nuclei, as long as the first excited state is not too low in energy (so that elastically scattered electrons can be separated from inelastic excitations).  In general, $R_n$ is easier to measure in lighter neutron rich nuclei.  This is because $R_n$ is smaller and so can be measured at higher momentum transfers where $A_{pv}$ is larger.   

Measuring $R_n$ in $^{48}$Ca is particularly attractive.  First, $^{48}$Ca has a higher experimental figure of merit than $^{208}$Pb.  Therefore a $^{48}$Ca measurement may take less beam time than for $^{208}$Pb.  Not only does $^{48}$Ca have a large neutron excess, it is also relatively light.  With only 48 nucleons, microscopic coupled cluster calculations \cite{coupledcluster}, or no core shell model calculations \cite{NCSM}, may be feasible for $^{48}$Ca that are presently not feasible for $^{208}$Pb.   Note that these microscopic calculations may have important contributions from three nucleon forces.  This will allow one to make microscopic predictions for the neutron density and relate a measured $R_n$ to three nucleon forces and in particular to very interesting three neutron forces.  

We end this section with a discussion of perhaps the ultimate neutron density measurement.  Measuring $A_{pv}$ for a range of momentum transfers $q$ allows one to directly determine the neutron density $\rho_n(r)$ as a function of $r$.   For $^{208}$Pb this may be extraordinarily difficult.  However, for $^{48}$Ca this may actually be feasible, although somewhat difficult and time consuming.  For example, one might be able to determine about 6 Fourier Bessel coefficients in an expansion of $\rho_n(r)$.  Note that this may require long runs and careful control of backgrounds and could be helped by using a large acceptance spectrometer.  Nevertheless, one should not underestimate the utility of having model independent determinations of both the neutron and proton densities as a function of $r$.  This will literally provide our picture of the neutrons and protons inside an atomic nucleus.

\section{Radiative corrections to PREX and QWEAK}
\label{Sec2}

\begin{figure}[h]
\center\includegraphics[width=4in]{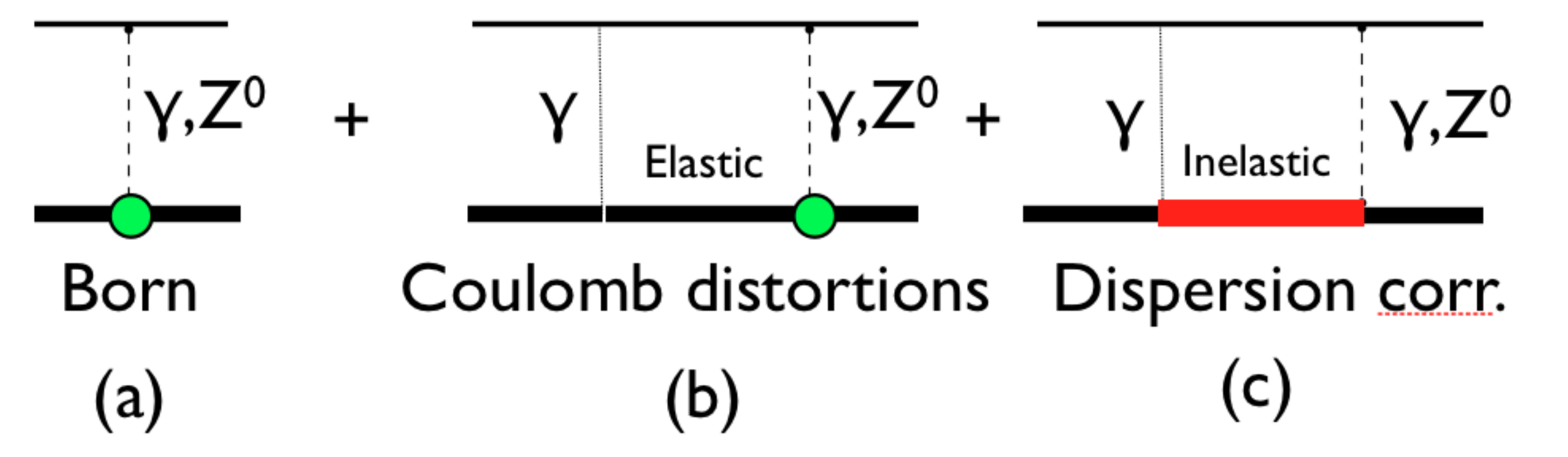}
\caption{\label{Fig2} Diagrams describing the electroweak scattering of an electron (top line) with a nucleus (bottom thick line), see text.}
\end{figure}

Radiative corrections involving additional photons are important for a variety of precision electromagnetic and electroweak experiments.  Coulomb distortion and dispersion corrections are illustrated in Fig. \ref{Fig2}.   Coulomb distortions, Fig. \ref{Fig2} (b), involve intermediate states where the nucleus remains in its ground state, while  dispersion corrections, Fig. \ref{Fig2} (c), involve intermediate states where the nucleus is excited.

The individual protons contribute coherently to Coulomb distortions, and as a result Fig. \ref{Fig2} (b) is order $Z\alpha$ compared to the Born term in Fig. \ref{Fig2} (a).  Here $Z$ is the atomic number and $\alpha$ is the fine structure constant.  Coulomb distortions are clearly important for PREX because of the large $Z=82$ of Pb.  Coulomb distortions can be included to all orders in $Z\alpha$ by numerically solving the Dirac Equation for an electron moving in the Coulomb potential of order 25 MeV and a weak axial vector potential of order 10 eV \cite{coulombdistortions}.  Coulomb distortions reduce $A_{pv}$ by about 30\% and are the largest known correction to the Born approximation.  However the uncertainty in this correction is very small because the charge density of $^{208}$Pb is accurately known.   The observed $A_{pv}$ is consistent with Coulomb distortion calculations and inconsistent with Born approximation predictions for any conceivable neutron density.   Therefore, PREX provides the first observation of Coulomb distortion effects in parity violating electron scattering.

In general the individual protons contribute incoherently to dispersion corrections.  Therefore Fig. \ref{Fig2} (c) is of order $\alpha$ rather than $Z\alpha$ and not obviously large for PREX.  There is some information on dispersion contributions to elastic electron scattering, see for example \cite{dispersion}.  However, dispersion corrections may be particularly important for parity violating electron scattering from the proton because the Born contribution in Fig. \ref{Fig2} (a) is small, of order $Q_W^p \approx 0.07$, where $Q_W^p$ is the small weak charge of the proton.  As a result dispersion corrections in Fig. \ref{Fig2} (c) can make a relative contribution of order $\alpha/Q_W^p\approx 10\%$.    Indeed Gorchtein first calculated a large dispersion correction to the weak charge of the proton that should be important for the interpretation of the QWEAK experiment \cite{G2009}.  Recently Gorchtein, Horowitz, and Ramsey-Musolf calculated this correction in more detail and estimated the remaining theoretical uncertainty \cite{G2011}.  About half of the correction involves nucleon resonance excited states while the remainder is from higher energy ``background'' excitations.  An important remaining uncertainty is the isospin decomposition of this background.  This is necessary to predict the electroweak response from purely electromagnetic data \cite{G2011}.  The dispersion correction $\Delta R$ predicted by \cite{G2011} to the weak charge of the proton is, 
\begin{equation}
\Delta R =[5.39\pm0.27\,(mod.\,avg.)\pm1.88\,(backgr.)^{+0.58}_{-0.49}\,(res.)\pm0.07\,(t-dep.)]\,10^{-3}.
\end{equation}
This is at the kinematics of the QWEAK experiment, $E=1.165$ GeV, and a squared momentum transfer of $t=-0.03$ GeV$^2$.  The first error involves the model used to fit electromagnetic data, while the second (largest) error involves uncertainties in high energy ``background'' contributions including their isospin dependence.  Finally smaller errors are quoted for the resonance contributions $(res.)$ and for the extrapolation from $t=0$, where the calculation was performed, to $t=-0.03$ GeV$^2$.  This correction $\Delta R$ is to be compared to the total weak charge of the proton $Q_W^p=0.0767\pm 0.0008\pm 0.0020$, for a relative contribution of $\Delta R/Q_W^p=7.6\pm 2.8 \%$, \cite{G2011}.  The correction is large compared to the desired 2\% statistical error for QWEAK.  Therefore it must be accurately included in the interpretation of experimental results.  The theoretical uncertainty can be reduced by measuring ``deep'' inelastic parity violating electron scattering at high excitation energies and moderate momentum transfers to constrain the isospin of background contributions.  Some other calculations of $\Delta R$ include \cite{A2011,B2011,R2011}.

At lower energies $\Delta R$ is both smaller and has a much smaller uncertainty.  At $E=0.180$ GeV, $t=0$ we calculate \cite{G2011},
$\Delta R =[1.32\pm0.05\,(mod.\,avg.)\pm0.27\,(backgr.)^{+0.11}_{-0.08}\,(res.)]\,10^{-3}$.
Therefore a low energy QWEAK like experiment, see for example \cite{MAINZ}, should be very cleanly interpretable in terms of the weak charge of the proton.

\acknowledgments
This work was done in collaboration with Mikhail Gorchtein and Shufang Ban and was supported in part by DOE grant DE-FG02-87ER40365.

\end{document}